\documentclass[twocolumn,showpacs,preprintnumbers,amsmath,amssymb,floatfix,prl,superscriptaddress]{revtex4-1}
\usepackage{epstopdf}
\usepackage[latin1]{inputenc} 
\usepackage{setspace}
\usepackage{graphicx}
\usepackage{graphics}
\usepackage{subfigure}
\usepackage{dcolumn}
\usepackage{bm}
\usepackage{color}
\usepackage{SIunits}
\usepackage[normalem]{ulem}
\usepackage{amsfonts}

\bibpunct{[}{]}{,}{n}{,}{,}

\begin{document}

\title{Controlling the velocity of ultrashort light pulses in vacuum \\through spatio-temporal couplings}

\author{A. Sainte-Marie}
\author{O. Gobert}
\author{F.Qu\'er\'e}
\affiliation{LIDYL, CEA, CNRS, Universit\'e Paris-Saclay, CEA Saclay, 91 191 Gif-sur-Yvette, France\\}


\date{\today}



\begin{abstract}
Due to their broad spectral width, ultrashort lasers provide new possibilities to shape light beams and control their properties, in particular through the use of spatio-temporal couplings. In this context, we present a theoretical investigation of the linear propagation of ultrashort laser beams that combine temporal chirp and a standard aberration known as longitudinal chromatism.
 When such beams are focused in a vacuum, or in a linear medium, the interplay of these two effects can be exploited to set the velocity of the resulting intensity peak to arbitrary values within the Rayleigh length, i.e. precisely where laser pulses are generally used. Such beams could find groundbreaking applications in the control of laser-matter interactions, in particular for laser-driven particle acceleration.
\end{abstract}

\maketitle

\section{Introduction}

Ultrashort laser pulses are being increasingly used in different fields of research and for a variety of applications \cite{laser}. This has stimulated the development of advanced tools to optimize and shape the properties of these beams. In the spatial domain, this is achieved using the same instruments as for continuous lasers, such as spatial light modulators \cite{SLM} or deformable mirrors \cite{DefMirror}. But what is obviously unique to ultrashort lasers is the possibility to shape light beams temporally \cite{PulseShaping}, by adjusting the relative phases of their multiple spectral components. Different types of programmable devices are now available to this end, such as acousto-optic dispersive filters \cite{TOURNOIS1997245, AOPDF}. 

One of the present frontiers in the shaping of ultrashort laser beams is the control of their spatio-temporal or spatio-spectral properties \textit{in a coupled manner}. A beam whose temporal or spectral properties depend on space -or vice versa- is said to exhibit spatio-temporal couplings (STC) \cite{Aktruk}. STC have long been considered as detrimental for applications of ultrashort lasers, as they increase the pulse duration and reduce the peak laser intensity at focus \cite{TERMITES}. 
Yet, there is increasing evidence for the advanced degrees of control that they can provide. They turn out to be extremely advantageous in nonlinear optics for elementary applications such as sum frequency generation \cite{Martinez, Gobert:14}, but also in more advanced schemes. For instance, a simple effect known as simultaneous spatio-temporal focusing has become widely exploited in nonlinear microscopy to increase the field of view without compromising depth resolution \cite{Oron:05, Zhu:05, Durfee}. Its close relative \cite{Durfee, JPhysB}, ultrafast wavefront rotation \cite{Akturk:05}, has enabled the generation of new ultrashort light sources called attosecond lighthouses \cite{fabien_light, pharexpCWE, pharexpGas}. More recently, theoretical studies have shown how a new type of linear non-diffracting beams could be produced by using ultrashort pulses where the $\textbf{k}$ vector of each plane wave component is correlated to its frequency in a specific manner \cite{Kondakci:16, Parker:16}. 

\begin{figure*}[t]
\centering \includegraphics[width=\linewidth]{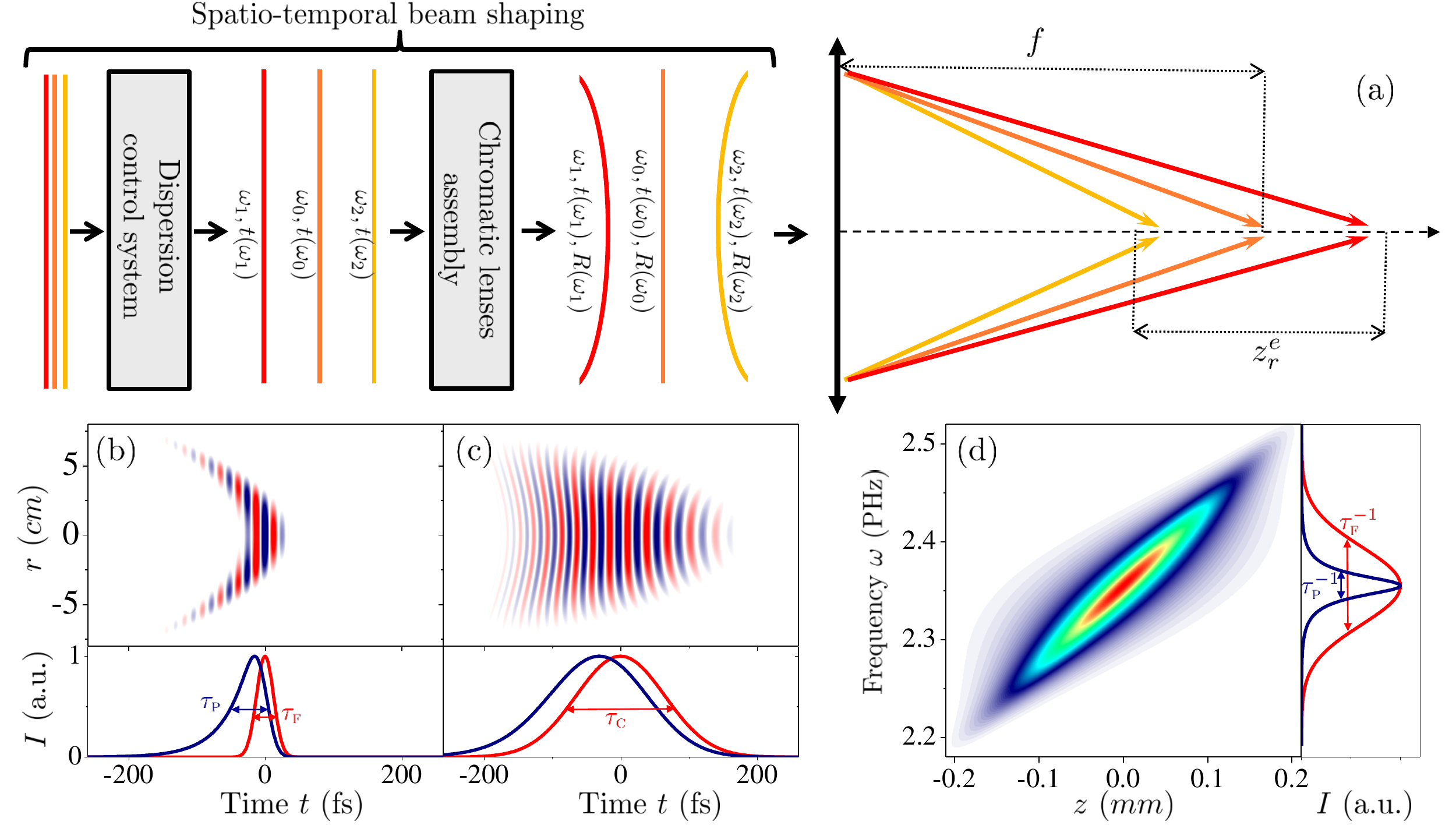}
\caption{\textbf{Generation and properties of CPLC beams} (a) A chromatic linear optical system induces a combination of temporal chirp and frequency-dependent wavefront curvature on an ultrashort beam. When this beam is subsequently focused, its different frequency components have their best focus at different $z$ (longitudinal chromatism). (b) Spatio-temporal field $E(\textbf{r},t)$ of the beam prior to focusing in a saturated color scale (negative field in blue, positive in red), in the presence of PFC/LC ($\alpha=3$ $fs/cm^2$ and beam waist $w_i=5$ $cm$ before focusing, corresponding to $\tau_{\textsc{p}}=\alpha w_i^2=75$ $fs$) and in the absence of chirp, for a local pulse duration $\tau_{\textsc{f}}=25$ $fs$.  The carrier frequency has been reduced for the sake of readability. Pulse front curvature is clearly observed. (c) Same plot, now in the presence of chirp ($\beta=-6380$ $fs^2$), i.e. for a CPLC beam. In (b) and (c), the lower panels show the local temporal intensity profile $\left|E(\textbf{r}=0,t)\right|^2$ at the center of the beam (red curve), and the spatially-integrated temporal intensity profile $\int d^2\textbf{r}\left|E(\textbf{r},t)\right|^2$ (blue curve). These profiles respectively have characteristic temporal widths of $\tau_{\textsc{f}}$ and $\tau_{\textsc{p}}$ in the absence of chirp, and $\tau_{\textsc{c}}\propto \left|\beta\right| \Delta \omega$ for strong chirps. (d) On-axis spectrum $\left|E'(z,\omega)\right|^2$ of the focused beam, as a function of the longitudinal coordinate $z$ along the extended Rayleigh length. The right panel compares the local spectrum at $z=0$ (blue curve, of characteristic width $1/\tau_{\textsc{p}}$) with the spatially-integrated spectrum $\int dz \left|E'(z, \omega)\right|^2$ (red curve, of characteristic width $1/\tau_{\textsc{f}}$).
} 
\vskip -0.5cm
\label{fig1}
\end{figure*}

Due to the technical difficulty of shaping laser beams in space-time or space-frequency, all experiments where STC have actually been put to use so far have been based on the lowest order couplings (i.e. linear with respect to both position and time/frequency, such as pulse front tilt or linear spatial chirp), which can be easily induced and varied using basic optical elements, e.g. prisms or gratings. One of the most elementary couplings of higher order, that typically results from propagation in lenses, is known as Pulse Front Curvature (PFC) in the near-field, and Longitudinal Chromatism (LC) in the far-field \cite{Bor1,Bor2,Bor3, Heuck2006, Rouyer:07, Neauport:07, Bahk:14}. To the best of our knowledge, there is at present no identified scheme to take advantage of this low-order coupling, despite its simplicity.

In this letter, we show how combining this well-known chromatic effect with temporal chirp (Fig. \ref{fig1}(a)) can provide advanced control on the velocity of the intensity peak formed by a laser pulse around its focus \cite{remark}, to arbitrary values either smaller or larger than $c$, and even up to regimes of apparent backward propagation or arbitrary longitudinal accelerations. For these Chirped Pulses with Longitudinal Chromatism (CPLC), the point of best focus moves along the pulse temporal envelope as the beam propagates through the Rayleigh length. This 'sliding focus' effect continuously reshapes the pulse temporal profile, resulting in a light burst with a tunable effective velocity $v$. We will only consider the case of light in a vacuum, but this effect equally applies to linear propagation in a medium, where the velocity of the pulse can thus be offset from the standard group velocity. 

CPLC beams open new possibilities for the control of laser-matter interaction, which come at the cost of a reduction in peak laser intensity relative to the focusing of a STC-free laser beam. Given the considerable powers now provided by ultrashort lasers \cite{PW}, this reduction should however remain tractable in many cases of interest. All laser beam parameters used in this article are typical of 100 TW-class femtosecond lasers used nowadays for laser-driven electron acceleration experiments \cite{Leemans}, for which controlling the light pulse velocity could be particularly relevant.

\section{Pulse front curvature and longitudinal chromatism}
\label{sec:PFC/LC}

\begin{figure*} []
\centering \includegraphics[width=\linewidth]{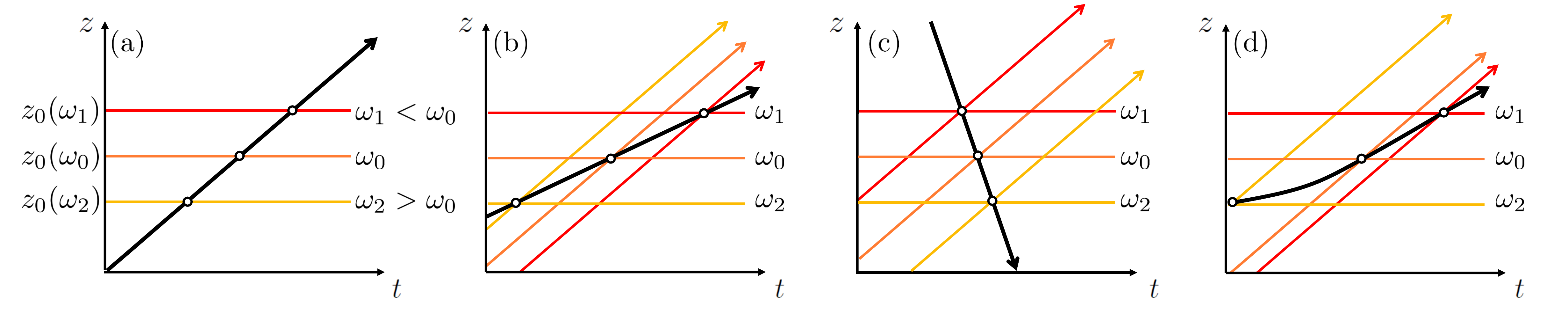}
\caption{\textbf{How CPLC enable adjustable light pulse velocities.}
  Due to PFC/LC, different frequencies of the pulse (here $\omega_1<\omega_0<\omega_2$) are focused at different longitudinal positions $z_0(\omega)$, indicated by the horizontal color lines. In (a), no chirp is applied, and the pulse intensity peak (black line with arrow) propagates at $c$. In (b), the frequencies that have their best focus at large $z$ are retarded in time, leading to an effective velocity smaller than $c$. In (c), these same frequencies are now advanced in time, in such a way that the intensity peak first appears at large $z$, and then moves in a direction opposite to the beam propagation direction -i.e. an apparent backward propagation within the extended Rayleigh length. In (d), a third-order spectral phase is applied to the beam, leading to a longitudinal acceleration of the pulse peak as it propagates. 
} 
\vskip -0.5cm
\label{fig2}
\end{figure*}

PFC and LC result from the propagation of an ultrashort beam through a chromatic lens, or a set of such lenses in a telescope for instance {\cite{Bor1,Bor2,Bor3}. After such a system, chromatism leads to a spatial curvature of the beam wavefront that varies with frequency (Fig. \ref{fig1}(a)). In the time domain, this corresponds to different curvatures for the wavefront and pulse front (also called energy front) -hence the name Pulse Front Curvature for the STC that affects the beam in the near-field (Fig. \ref{fig1}(b)). As this beam gets focused (either by the chromatic system itself, or by a subsequent achromatic focusing optic), the different frequency components have their best focus at different longitudinal positions (Fig. \ref{fig1}(d)), due to their different wavefront curvatures - hence the name Longitudinal Chromatism for the spatio-spectral coupling affecting such a beam in the far-field. PFC and LC can thus be viewed as two facets of the same aberration, occurring at different stages of the beam propagation: in the following, we will therefore refer to this aberration as PFC/LC. We will now express this qualitative reasoning in mathematical terms.

The spatio-temporal $E$-field of a collimated laser beam with PFC can be written, for a given propagation coordinate $z$, as $E(\textbf{r},t)=\exp({i\omega_0 t}) A_0(t-\alpha \left|\textbf{r}\right|^2)$, with $\textbf{r}$ the spatial coordinate transverse to the beam propagation direction, $\omega_0$ the pulse carrier frequency, and $A_0$ its complex envelope. $\alpha$ is the parameter that quantifies the strength of the PFC: at the radial position $r=w_i$ ($w_i$ waist of the collimated beam), the pulse envelope is delayed by $\tau_{\textsc{p}}=\alpha w_i^2$ with respect to the one at $r=0$. 
Fourier-transforming this expression with respect to time provides the spatio-spectral field, $E'(\textbf{r},\omega)=\exp(- i \alpha \delta \omega\left|\textbf{r}\right|^2)A'_0(\delta \omega)$, with $\delta \omega = \omega- \omega _0$ the frequency offset with respect to $\omega_0$, and $A'_0$ the Fourier transform of $A_0$. Writing the spatial phase at frequency $\omega$ as $\varphi(\textbf{r},\omega)=-\omega \left|\textbf{r}\right|^2/2cR(\omega)$, with $1/R(\omega)$ the frequency-dependent wavefront curvature, we get $1/R(\omega)= 2 c \alpha \delta \omega/\omega$. 

After a perfect focusing element of focal length $f$, the wavefront curvature becomes $1/R'(\omega)=1/R(\omega)-1/f$. If the beam impinging the lens is quasi-collimated ($R(\omega)\gg f$), this leads to $R'(\omega)\approx -(f+f^2/R(\omega))$. For physically sound cases, the position $z_0(\omega)$ of the best focus for frequency $\omega$ is at a distance $-R'(\omega)$ from the focusing optic. Taking the convention $z_0(\omega_0)=0$ and assuming for simplicity that $\delta \omega \ll \omega_0$, we thus get:
\begin{equation}
z_0(\omega)=  2f^2 c \alpha \delta \omega/\omega_0
\label{LC}
\end{equation}
 This is longitudinal chromatism, i.e. a linear drift of the best focus position with frequency. This relationship can also be written as $z_0(\omega)= \tau_{\textsc{p}} z_r \delta \omega$, where $z_r=\lambda_0 f^2/\pi w_i^2$ is the Rayleigh length of the focused beam formed by frequency $\omega_0=2\pi c/\lambda_0$. PFC/LC thus leads to an extended \textsl{frequency-integrated }Rayleigh length $z_r^e$, with $z_r^e/z_r = \tau_{\textsc{p}} \Delta \omega \propto \tau_{\textsc{p}}/\tau_{\textsc{f}}$ in the limit of large $\tau_{\textsc{p}}$, where $\Delta \omega$ is the spectral width of the pulse, and $\tau_{\textsc{f}}$ its Fourier-transform limited duration. This increased Rayleigh range can in itself be of interest for some applications.

\section{Control of the intensity peak velocity using temporal chirp}

The key idea of the present paper is simple and intuitive (Fig. \ref{fig2}): 
with the frequency components of the ultrashort pulse focused at different longitudinal positions $z_0(\omega)$ due to PFC/LC, the arrival time of light at each $z$ can be controlled simply by adjusting the relative timing of these frequencies in the pulse prior to focusing (Fig. \ref{fig1}(c)), using standard temporal pulse shaping techniques. This provides a programmable control on the pulse propagation velocity along the Rayleigh length. The qualitative sketches of Fig. \ref{fig2} illustrate how this idea enables arbitrary pulse velocities around focus, such as $v<c$ (panel b), $v<0$ (panel c), or longitudinally accelerating beams (panel d). 

We will now confirm this idea by deriving a simple analytical formula for the effective propagation velocity of the pulse around focus, in the case of linear temporal chirp (i.e. quadratic spectral phase), corresponding to the application, in addition to PFC/LC, of a spatially-homogeneous spectral phase $-\beta \delta \omega ^2/2$. Large values of $\beta$ ($\beta \gg \tau_{\textsc{f}}^2$) lead to a well-defined linear relationship between frequency and arrival time, such that frequency $ \omega$ arrives at position $z$ at time:
\begin{equation}
t(\omega)\equiv -\frac{\partial \varphi (z,\omega)}{\partial \omega}=z/c+\beta \delta \omega
\label{Chirp}
\end{equation}
The propagation velocity $v$ of the intensity peak formed in the extended Rayleigh length can now be easily calculated by applying Eq. (\ref{Chirp}) to the specific position $z=z_0(\omega)$, and combining the resulting equation with Eq. (\ref{LC}) to eliminate $\delta \omega$ and obtain a direct relationship between $z_0(\omega)$ and $t_0(\omega)$. This leads to $z_0(\omega)=v t_0(\omega)$ with:
\begin{equation}
\frac{v}{c}=\frac{1}{1+ (\omega_0/2  f^2) \times (\beta/\alpha)}
\label{velocity}
\end{equation}
As expected from the qualitative sketches of Fig. \ref{fig2}, depending on the value of $\beta/\alpha$ -the ratio  of the chirp and PFC/LC parameters, $\left|v\right|$ can be either larger or smaller than $c$ (see also black curve in Fig. \ref{fig5}(a)). It becomes infinite (i.e. the maximum intensity occurs at the same instant at all $z$) for a spectral chirp of $\beta_0=-2 f^2 \alpha / \omega_0$ 
, and then negative as $\left|\beta\right|$ is further increased. Physically, such negative velocities corresponds to a regime where the frequencies focused at large $z$ have a temporal advance such that they reach their best focus before the frequencies focused at smaller $z$ reach theirs. This results in an intensity peak that effectively propagates \textit{backward} along the extended Rayleigh range $z_r^e$ (Fig. \ref{fig2}(c)).

\section{Theoretical description}

The properties of CPLC can be calculated more formally, under the assumption that each spectral component of the pulse has a Gaussian spatial profile. According to the previous discussion, after a focusing optic, the beam consists of a superposition of Gaussian beams of different frequencies $\omega$, with their waist located at different positions $z_0(\omega)=\tau_{\textsc{p}} z_r \delta \omega$. Using the known analytical expression of Gaussian beams, and restricting the discussion to on-axis positions ($\textbf{r}=0$) for simplicity, the spatio-spectral field $E'(z,\omega)$ writes:
\begin{equation}
E'(z,\omega)=  \frac{\left|A'_0(\delta \omega)\right|}{\left[1+(z-\tau_{\textsc{p}} z_r \delta \omega)^2/z_r^2\right]^{1/2}} e^{i\varphi(z,\omega)}
\label{Amplitude}
\end{equation}
with
\begin{equation}
\varphi(z,\omega)=\arctan\left(\frac{z-\tau_{\textsc{p}} z_r \delta \omega}{z_r}\right)-\frac{\beta}{2}\delta \omega^2 - \frac{\omega}{c}z
\label{Phase}
\end{equation}
The denominator in Eq. (\ref{Amplitude}) accounts for the evolution of the on-axis field amplitude resulting from focusing, and the first term in Eq. (\ref{Phase}) originates from the Gouy phase which, in the presence of LC, becomes a function of both $z$ and $\omega$. 

To understand the \textit{spectral} properties of these beams, we momentarily consider the limit of large PFC/LC ($\tau_{\textsc{p}} \gg \tau_{\textsc{f}}\propto1/\Delta \omega$), such that the variations of the numerator in Eq. (\ref{Amplitude}) are much slower than those of the denominator. Within the extended Rayleigh range $z_r^e$, the argument of $A'_0(\delta \omega)$ in Eq. (\ref{Amplitude}) can then be replaced by $\delta \omega\approx z/\tau_{\textsc{p}} z_r=\Delta \omega z/z_r^e$, leading to the following expression of the spectral intensity:
\begin{equation}
\left|E'(z,\omega)\right|^2 \propto \frac{\left|A_0'(\Delta \omega z/ z_r^e)\right|^2}{\left[1+\tau_{\textsc{p}}^2(\delta \omega-\Delta \omega z/ z_r^e)^2\right]} 
\label{LargePFC}
\end{equation}
This reveals a key feature of beams with PFC/LC, illustrated in Fig. \ref{fig1}(d): for large enough PFC/LC, the \textit{local} spectrum within the extended Rayleigh range is a Lorentzian function, of width $1/\tau_{\textsc{p}}$ smaller than the initial spectral width $\Delta \omega$ of the ultrashort pulse, with a central frequency $\omega_0+ \Delta \omega z/ z_r^e$ that depends linearly on $z$. Moreover, the longitudinal on-axis fluence distribution of the focused beam no longer has the Lorentzian profile of standard Gaussian beams: this profile is instead imposed by the spectrum $\left|A_0'\right|^{2}$ of the initial ultrashort pulse, which is spatially-stretched along the length $z_r^e$.

\begin{figure*} []
\centering \includegraphics[width=\linewidth]{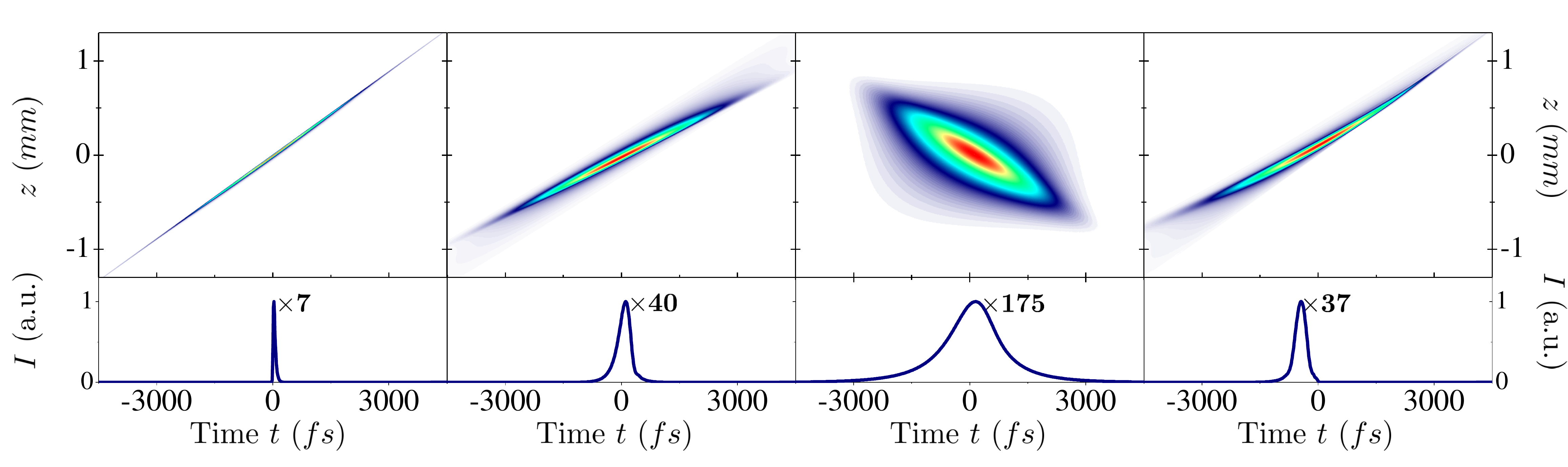}
\caption{\textbf{Simulations of CPLC of different propagation velocities.} 
Simulated on-axis spatio-temporal intensity profile $I(z,t)$ of different CPLC beams along the extended Rayleigh length, for a fixed PFC/LC parameter $\alpha=3$ $fs/cm^2$ and different spectral phases. The beam parameters are $\lambda_0=800$ $nm$, $\tau_{\textsc{f}}=25$  $fs$, $w_i=5$ $cm$, focused by an optic of focal length $f=1$ $m$ (i.e. $f/20 $ focusing). Panel (a) corresponds to a reference case without chirp, where the pulse envelope propagates at $c$. In (b) and (c), a linear chirp has been applied ($\beta=12230$ $fs^2$ and $\beta=-53420$ $fs^2$), that respectively lead to the propagation regimes of Fig. \ref{fig2}(b) and (c), with $v=0.7\:c$ in (b) and $v=-c$ in (c). In (d), a third-order spectral phase is applied to the beam, leading to a longitudinal acceleration of the pulse as it propagates, as illustrated in Fig. \ref{fig2}(d). In each case, the lower plots show the pulse temporal intensity profile at $z=0$, $I(0,t)$. By convention, $I=1$ corresponds to peak intensity obtained at best focus for the STC-free unchirped beam. For comparison with this reference case, all curves have been multiplied by the numerical factors indicated next to the curves, which thus correspond to the inverse of the intensity reduction factor $\epsilon$ resulting from the combination of PFC/LC and chirp.
} 
\vskip -0.5cm
\label{fig3}
\end{figure*}

The spatio-temporal field $E(z,t)$ of the beam is the inverse Fourier transform of Eq. (\ref{Amplitude}), $E(z,t) = \int \frac{d\omega}{2\pi} E(z,\omega) e^{i \omega t}$. This integral can be estimated analytically in the limit of large chirp $\beta$, using the stationary phase method \cite{math}. This requires finding the frequency $\omega$ such that $\partial (\varphi(z,\omega)+\omega t)/\partial \omega=0$. This is easily solved provided the Gouy phase is neglected compared to the quadratic spectral phase $-\beta \delta \omega^2/2$ in Eq. (\ref{Phase}), and leads to $\delta \omega=\left[t-z/c\right]/\beta \equiv t'/\beta$, which is identical to Eq.(\ref{Chirp}). $E(z,t)$ is then obtained essentially by inserting this expression of $\delta \omega$ in the integrand $E'(z,\omega) e^{i \omega t}$ (Eq. \ref{Amplitude} and \ref{Phase}), leading to:
\begin{equation}
E(z,t) \propto  \frac{\left|A'_0(t'/\beta)\right|}{\left[ 1+ (t''/\tau_{\textsc{e}})^2  \right]^{1/2}} e^{i\phi(z,t)}
\label{profiltemporel}
\end{equation}
with
\begin{eqnarray} 
t''&=&t'- \frac{\beta}{\tau_p z_r} z \\
&=&t-\frac{z}{v}
\end{eqnarray} 
with $v$ given by Eq. (\ref{velocity}), $\tau_{\textsc{e}}=\beta/\tau_{\textsc{p}}$, and
\begin{equation}
\phi(z,t)= (\omega_0+\frac{t'}{2\beta})t' - \arctan \left(t''/\tau_{\textsc{e}}\right)
\label{phasetemporelle}
\end{equation}
Equation (\ref{profiltemporel}) shows that the pulse temporal amplitude profile is the product of two terms  $p_1(z,t)$ and  $p_2(z,t)$.  (i) $p_1(z,t)=\left|A'_0(t'/\beta)\right|$ corresponds to the undistorted temporal envelope of the initial pulse, i.e. it is a chirped pulse of spectral width $\Delta \omega$ and duration $\tau_{\textsc{c}} \propto \left|\beta\right| \Delta \omega$, that propagates at $c$.  (ii) $p_2(z,t)=\left[ 1+ (t''/\tau_{\textsc{e}})^2\right]^{-1/2}$ corresponds to the envelope of a chirped pulse of spectral width $1/\tau_{\textsc{p}}$ and duration $\tau_{\textsc{e}}=\beta /\tau_{\textsc{p}}$, and it propagates at a velocity $v$ given by Eq. (\ref{velocity}). This term reshapes the initial pulse temporal envelope $p_1(z,t)$ as the beam travels along the Rayleigh length, thus affecting the effective propagation velocity of the peak of $\left|E(z,t)\right|$: this is the origin of the sliding focus effect. 

This expression further simplifies by again considering the limit of large PFC/LC, in addition to large chirp. In this limit, the temporal width of $p_2(z,t)$ is thus much smaller than the one of $p_1(z,t)$, because its smaller spectral width $1/\tau_p \ll \Delta \omega$ (Fig. \ref{fig1}(d)) leads to a weaker temporal stretching by the applied chirp. We can then replace $p_1(z,t)$ in Eq. (\ref{profiltemporel}) by its value at time $t''=0$ where $p_2(z,t)$ reaches its peak value. This directly leads to:
\begin{equation}
I(z,t) \equiv \left|E(z,t)\right|^2 \propto \frac{\left|A'_0(\Delta \omega z/ z_r^e)\right|^2}{\left[ 1+ (t''/\tau_{\textsc{e}})^2  \right]}
\label{profiltemporel2}
\end{equation}
In this regime, the pulse shape remains the same all along its propagation over the Rayleigh length at a velocity $v$. Note that for large chirps (such that the Gouy phase can be neglected), according to Eq. (\ref{phasetemporelle}), the phase velocity $v_{\varphi}$ is almost unaffected, $v_{\varphi} \approx c$, so that the carrier wave slips with respect to the pulse envelope, akin to a pulse propagating in a dispersive medium.

\section{Numerical simulations}

We now turn to numerical simulations of the pulse propagation, which have three main interests: (i) Showing that the sliding focus effect does not rely on the assumptions of large chirp or PFC/LC, which are only useful for analytical calculations. (ii) Demonstrating the possibility of inducing longitudinal acceleration of the pulse, by applying spectral phases of higher orders. (iii) Calculating the full spatio-temporal profile $E(z,\textbf{r},t)$ of the beam, to study how this profile is affected by the applied shaping and determine the resulting loss in peak intensity compared to a perfect STC-free beam. For these simulations, the field profile in the $(\textbf{r},\omega)$ space is calculated at different $z$ by numerically summing Gaussian beams with a frequency-dependent waist position (Eq. (\ref{Amplitude}-\ref{Phase})), with a  pulse spectrum now assumed to be Gaussian, $A'_0(\delta \omega)=e^{-\delta \omega^2/\Delta \omega^2}$. The spatio-temporal field $E(z,\textbf{r},t)$ is then calculated by a numerical Fourier-transform with respect to $\omega$.

Figure 4 displays plots of the simulated on-axis temporal intensity profiles $I(z,t)$ of CPLC beams along the extended Rayleigh length, for a fixed PFC/LC ($\alpha=3$ $fs.cm^{-2}$) and different applied spectral phases. For these four cases, simulations of the full spatio-temporal intensity $\left|E(z,\textbf{r},t)\right|^2$ profile are also provided in movie M1 of the supplementary material. Panel (a) is the reference case without chirp ($\beta=0$), such that the pulse propagates at $c$, the usual light velocity in vacuum. In this case, due to PFC/LC, $I(z,t)$ has a half-exponential temporal profile, as already demonstrated and clearly explained in \cite{Bor3}. In panels (b) and (c), two different linear chirps have been applied, that respectively lead to propagation velocities $v=0.7 c$ and $v=-c$. These two examples corresponds to the two regimes qualitatively sketched in Fig. \ref{fig2}(b) and (c). 

In Fig. \ref{fig3}(d), we explore a case of velocity control through a more complex spectral phase. While such cases cannot be easily handled analytically, a simple formula relating the pulse group delay dispersion $\partial^2 \varphi/\partial \omega^2$ to the desired velocity evolution $v(z)$ can still be obtained, by generalizing Eq. (\ref{Chirp}) to the case of arbitrary spectral phases $\varphi(\omega)$. This leads to:
\begin{equation}
\frac{\partial^2 \varphi}{\partial \omega^2}=\tau_{\textsc{p}} z_r\left(\frac{1}{v(z)}-\frac{1}{c}\right)
\label{casgeneral}
\end{equation} 
where $v(z)$ is related to $\omega$ through $z=z_r \tau_p \delta \omega$. The case shown in Fig. \ref{fig3}(d) is inspired from Fig. \ref{fig2}(d): in addition to PFC/LC, a quadratic group delay (cubic spectral phase), estimated using a Taylor expansion of Eq. (\ref{casgeneral}), has been applied on the pulse prior to focusing. This results in a progressive acceleration of the pulse, from $v\approx 0.6 c$ to $v=c$, along the extended Rayleigh length. Such accelerating pulses might prove useful for particle acceleration in the wake excited by ultraintense lasers in low-density plasmas, provided nonlinear propagation effects do not alter too much the sliding focus effect. The lower initial velocity could facilitate particle injection in this wake in the early stage of the process. Later in the interaction, particle deceleration by dephasing could be avoided by forcing the pulse velocity in the plasma to $c$, instead of the group velocity $v_g<c$.

\begin{figure*}[]
\centering \includegraphics[width=\linewidth]{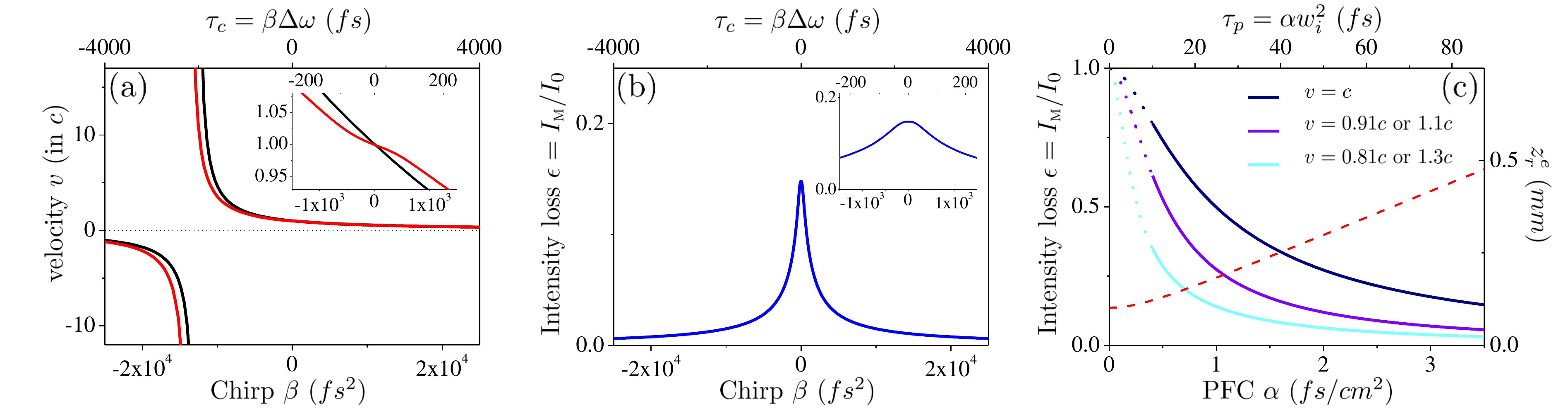}
\caption{\textbf{Quantitative properties of CPLC beams}, for $\alpha=3$ $fs/cm^2$, $\lambda_0=800$ $nm$, $\tau_{\textsc{f}}=25$ $fs$, $w_i=5$ $cm$ and $f=1$ $m$ as in Fig. \ref{fig3}. (a) Propagation velocity $v$ of the intensity peak of a CPLC beam around focus, as a function of chirp $\beta$. The black line corresponds to the prediction of Eq. (\ref{velocity}), while the red one shows the velocity deduced from numerical simulations.  
(b) Corresponding peak intensity reduction factor $\epsilon=I_{\textsc{m}}/I_0$ as a function of $\beta$. In both panels, the insets show zooms on these curves for small values of $\beta$. (c) Peak intensity reduction factor $\epsilon$ (full lines) as a function of $\alpha$, for fixed propagation velocities $v=c$, $v=1.1c$ or $0.91c$, and $v=1.3c$ or $0.81c$, corresponding to $\beta/\alpha=0$, $\pm 7.71.10^{2}$ $fs.cm^2$ and $\pm 1.95.10^{3}$ $fs.cm^2$. The dotted parts of the curves correspond to a range where the calculation no longer makes sense, because PFC/LC becomes too weak. The red dashed line shows the evolution of the extended Rayleigh length $z_r^e$ with $\alpha$, obtained from numerical simulations. 
} 
\vskip -0.5cm
\label{fig5}
\end{figure*}

\section{Superluminal velocities and causality}

For a deeper insight into the behavior of CPLC beams, movie M2 of the supplementary material shows the evolution of $\left|E(z,\textbf{r},t)\right|^2$ around and all along the extended Rayleigh range, for the case of Fig. \ref{fig3}(b) only, now with an evolving color scale normalized to the maximum of each image, so that the beam can still be seen as it defocuses. 
Out of the extended Rayleigh length, the light bullet propagates at $c$. When approaching focus, the pulse envelope starts being reshaped by the differential focusing of the pulse temporally-stretched frequencies. This leads to a pulse shortening, and the shortened light peak moves along the initial pulse envelope as the beam propagates (sliding focus effect), resulting in an effective peak velocity $v \neq c$. As the pulse defocuses, it recovers its initial envelope, and again propagates at $c$. Overall, no delay has been accumulated by the pulse along this path compared to a STC-free beam propagating at $c$. 

This analysis makes it clear -insofar as that may be necessary- that the superluminal velocities that can be obtained with CPLC do not violate causality and the postulate of relativity on the maximum possible speed of signal transmission. Like in several other physical processes \cite{Reviewc1,Reviewc2}, the local 'anomalous' velocity of the intensity peak results from the distortion of the pulse envelope as it propagates, and the peaks occurring at different $z$ are not causally connected. More generally, since the seminal work of Brillouin and Sommerfeld \cite{Brillouin}, it is well-known that the propagation of an intensity peak at a velocity $v>c$ does not imply that a signal can actually be transmitted at this velocity \cite{Reviewc1}. CPLC beams provide a new and instructive illustration of this general idea for propagation in vacuum: in Eq. (\ref{profiltemporel}), only the $p_1(z,t)$ term -propagating at $c$- can actually carry a signal, while the motion of the intensity peak is mainly determined by the $p_2(z,t)$ term -propagating at $v$. 

\section{Choice of the control parameters}

We finally discuss the essential point of the choice and optimization of the control parameters $\alpha$ and $\beta$. The use of CPLC to control the pulse velocity has two main drawbacks, which are (i) an increase in pulse duration at focus and (ii) a decrease in peak intensity, compared to those of an unchirped STC-free laser pulse \cite{Heuck2006}. The larger $\alpha$ and $\beta$, the stronger these effects. For applications where intensity and pulse duration are critical, $\alpha$ and $\beta$ thus need to be carefully chosen, such that one gets the required control on the pulse velocity while limiting the degradation of the pulse properties. 

For a fixed PFC/LC parameter $\alpha$, the value of the chirp parameter $\beta$ is imposed by the pulse velocity $v$ that one is aiming at. The red line in Fig. \ref{fig5}(a) shows the pulse velocity deduced from numerical calculations as a function of $\beta$, for $\alpha=3$ $fs/cm^2$. This curve is in good qualitative agreement with the prediction of Eq. (\ref{velocity}) (black curve), showing that this simple expression of the pulse velocity, derived under the approximation of large chirp, actually provides a good estimate of the light peak velocity in most cases. Fig. \ref{fig5}(b) shows the corresponding evolution of the reduction factor $\epsilon=I_{\textsc{m}}/I_0$ of the peak intensity $I_{\textsc{m}}$ of the CPLC beam (occurring at $z\approx0$ and $\textbf{r}=0$), relative to the one $I_0$ obtained with $\alpha=\beta=0$ for a beam of same energy and same spectrum. For $\beta=0$, $\epsilon$  is already smaller than 1 due to PFC/LC, and it then continuously decreases as $\beta$ gets larger. Larger deviations from $c$ thus imply stronger degradations of the peak intensity and pulse duration of the focused beam.  The insets of Fig. \ref{fig5}(a) and (b) show that for physically sound parameters, the reduction in peak intensity is less than ten fold when $v$ deviates from $c$ (or $v_g$ in a medium) by a few percent only, which typically correspond to the deviations needed to avoid dephasing in laser wakefield acceleration.  

The choice of $\alpha$ is rather the result of a compromise. For a given focusing geometry, a larger value of $\alpha$ leads to a larger extended Rayleigh length $z_r^e \propto \alpha$ (red curve in Fig. \ref{fig5}(c)) over which the pulse velocity can be controlled. Increasing $\alpha$ also enables a more sophisticated control of the pulse propagation, using more complex spectral phases: qualitatively, the number of degrees of freedom $N$ varies as $N\approx\Delta \omega \tau_{\textsc{p}} \propto \alpha$, the ratio of the total and local spectral widths around focus. As an illustration, using $N \approx 60$ and an oscillating spectral phase, one gets the ultrashort pulse displayed in Fig. \ref{fig4}, performing a succession of $\approx 5$ longitudinal accelerations and decelerations between $v=c$ and $v=0.3c$. 

On the other hand, increasing $\alpha$ leads to a stronger reduction in peak intensity and increase in pulse duration, even in the absence of chirp. To reach a given velocity $v$, a larger value of the chirp $\beta$ is then needed, resulting in an even larger reduction in $I_{\textsc{m}}$. To illustrate this fact, Fig. \ref{fig5}(c) shows the evolution of $\epsilon$ as a function of $\alpha$, for three values of the ratio $\left| \beta/\alpha \right|$, corresponding to different velocities $v$ either larger or smaller than $c$. Note that the influence of the focal length $f$ of the focusing optic is qualitatively similar to that of $\alpha$ (Eq. (\ref{velocity})): for a given beam diameter $w_i$, increasing $f$ leads to a larger value of $z_r^e$, but larger chirps $\beta$ are then needed to achieve a given velocity $v \neq c$, resulting in stronger reductions in peak intensity.

\section{Conclusion and perspectives}

In conclusion, the sliding focus effect occurring on CPLC beams provide a new powerful approach to change the propagation velocity of light pulses over finite distances in vacuum as well as in linear media. Once PFC/LC has been applied on a beam, e.g. using a set of chromatic lenses or more advanced new shaping techniques \cite{Sun:15}, the pulse velocity around the beam focus can be changed in a programmable manner, thanks to the optical devices now routinely used to tailor the spectral phase of ultrashort light pulses \cite{TOURNOIS1997245, AOPDF}. This control scheme is in principle applicable to any other type of waves, be they classical or quantum \cite{Giovannini857}.

This new type of beam should prove useful for the control of laser-matter interaction effects occurring over extended distances, and where light intensity is the relevant physical parameter. For instance, the possibility to generate intense 'slow' ($v \ll c$) pulses could facilitate laser-driven acceleration of heavy particles, such as ions \cite{RevModPhys.85.751}.  Fine tuning of the velocity around $c$ could be beneficial for laser wakefield acceleration of relativistic electrons \cite{Leemans}, or to avoid velocity mismatch effects in experiments involving the propagation of multiple pulses of different frequencies in a dispersive medium \cite{2015NaPho...9..817D}. Backward propagation of the intensity peak could make it possible to revert the propagation of secondary emissions that normally only occur in the forward direction of the driving laser beam. In many of these applications though, the beam propagation in the medium is actually nonlinear, which might alter the beam spatio-temporal properties and hence the velocity of the CPLC intensity peak . Further studies will be required to determine the influence of these effects. 

\begin{figure} []
\centering \includegraphics[width=0.9\linewidth]{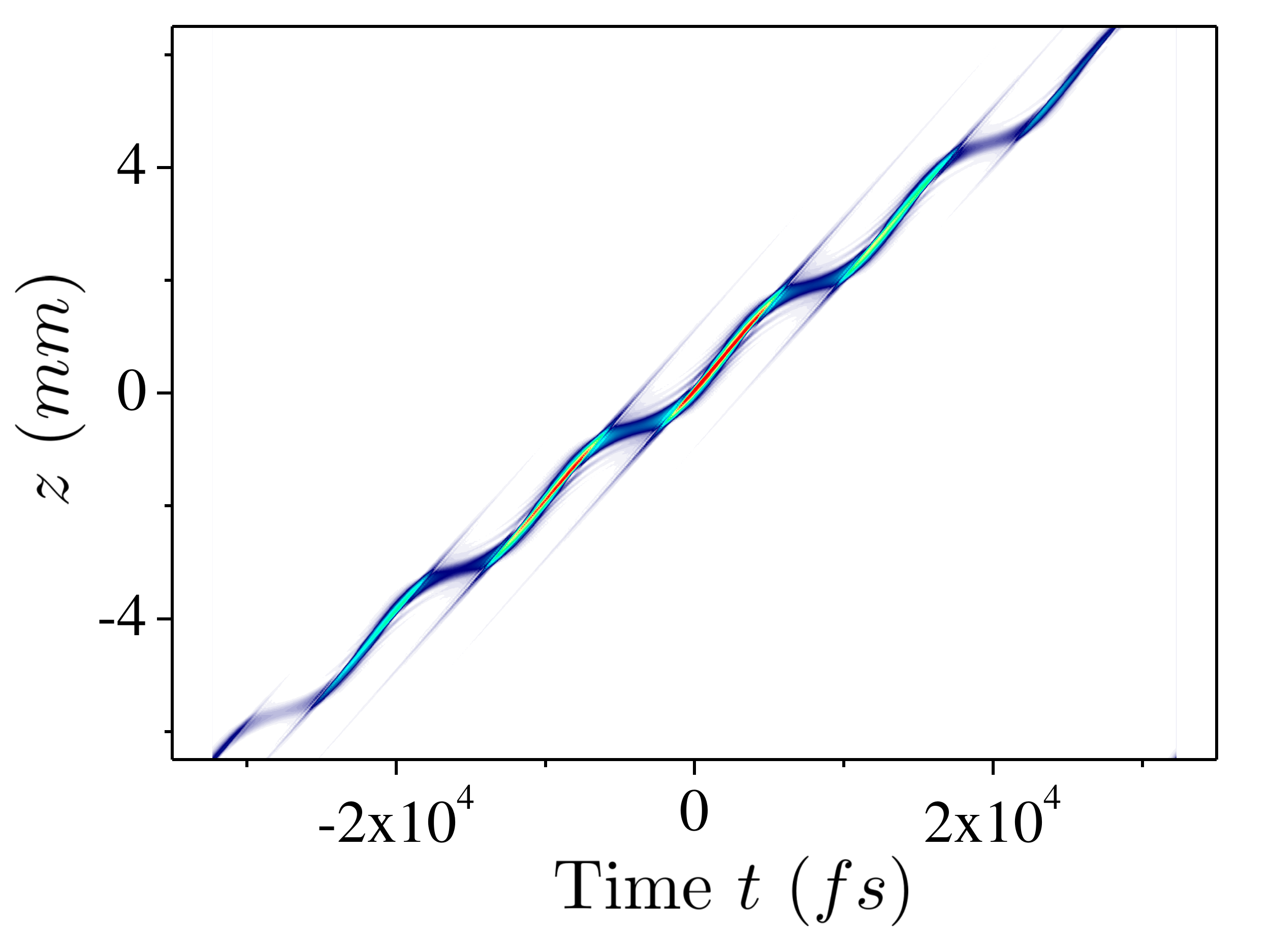}
\caption{\textbf{Pulse with an oscillating propagation velocity produced using CPLC with a high-order spectral phase.} 
Simulated on-axis spatio-temporal intensity profile $I(z,t)$, along the extended Rayleigh length, of a CPLC beam with an oscillating spectral phase, leading to a velocity $v$ oscillating between $c$ and 0.$3c$ . Compared to Fig. \ref{fig3}, the PFC/LC parameter has also been increased to $\alpha=30$ $fs/cm^2$ ($\tau_{\textsc{p}}=750$ $fs$), while all other beam parameters are identical. 
} 
\vskip -0.5cm
\label{fig4}
\end{figure}

\section*{Funding Information}
European Research Council (ERC) (Adv. grant 694596).

\section*{Acknowledgments}

F.Q. gratefully acknowledges Dr. C\'edric Thaury for the initial discussions on the influence of the combination of PFC/LC and temporal chirp on laser wakefield acceleration, that triggered this work. 



\section*{References}



\begin{thebibliography}{38}
\bibitem [{rem()}]{remark}%
  \BibitemOpen
  \href@noop {} {}\bibinfo {note} {This velocity might be considered as the
  local group velocity of the pulse within the Rayleigh length. However the
  usual concept of group velocity, well-suited to the analysis of wave
  propagation in dispersive media, does not come into play in the present
  theoretical analysis. We have therefore chosen not to use this term
  here.}\BibitemShut {Stop}%
\end{thebibliography}

%

}{}


\end{document}